\pdfoutput=1
\documentclass{article}

\usepackage[english]{babel}

\usepackage[letterpaper,top=2cm,bottom=2cm,left=3cm,right=3cm,marginparwidth=1.75cm]{geometry}
\usepackage[round]{natbib}

\usepackage{amsmath}
\usepackage{graphicx}
\usepackage{tikz}
\usetikzlibrary{shapes.geometric,arrows.meta,positioning,calc,fit,backgrounds,decorations.pathreplacing}
\usepackage{xcolor}

\definecolor{gdmblue}{RGB}{26,115,232}
\definecolor{gdmgreen}{RGB}{24,128,56}
\definecolor{gdmyellow}{RGB}{251,188,5}
\definecolor{gdmred}{RGB}{234,67,53}
\definecolor{openminedpurple}{RGB}{114,9,183}
\definecolor{enclavegray}{RGB}{248,249,250}
\definecolor{darkgray}{RGB}{60,64,67}
\definecolor{lightgray}{RGB}{218,220,224}
\definecolor{stepbg}{RGB}{232,240,254}
\definecolor{attestbg}{RGB}{243,232,255}

\usepackage{caption}
\usepackage{subcaption}
\usepackage[colorlinks=true, allcolors=blue]{hyperref}

\title{Double Blind Evals: Resolving the Dual Confidentiality Dilemma in AI Safety Auditing}

\author{Andrew Trask\textsuperscript{2,4} \and Sol Messing\textsuperscript{2}
\and Vai Pahwa\textsuperscript{2} \and Pegah Maham\textsuperscript{2} \and Rene Kolga\textsuperscript{2} \and Austin Frantz\textsuperscript{2} \and Andrew Tash\textsuperscript{2} \and Kate Thomas\textsuperscript{2}
\and Sean McGregor\textsuperscript{1} \and George Balston\textsuperscript{1} \and Patricia Paskov\textsuperscript{1} \and Miles Brundage\textsuperscript{1} \and Akriti Vij\textsuperscript{3} \and Bennett Hillenbrand\textsuperscript{5} \and Alexandros Karargyris\textsuperscript{5}
\and Tin Acosta\textsuperscript{2} \and Joel Fenster\textsuperscript{2} \and Madeleine Elish\textsuperscript{2} \and Rasmi Elasmar\textsuperscript{2} \and Myriam Khan\textsuperscript{2} \and Koen van der Veen\textsuperscript{4} \and Rasswanth S\textsuperscript{4} \and Sameer Wagh\textsuperscript{4} \and Stephen Gabriel\textsuperscript{4} \and Pedro Werneck\textsuperscript{4} \and Lacey Strahm\textsuperscript{4} \and Kevin McDonough\textsuperscript{4} \and Ronnie Falcon\textsuperscript{4}
\and Kristian Lum\textsuperscript{2} \and William Isaac\textsuperscript{2}}

\date{
  \textsuperscript{1}AVERI \quad
  \textsuperscript{2}Google \quad
  \textsuperscript{3}Singapore AISI \quad
  \textsuperscript{4}OpenMined \quad
  \textsuperscript{5}MLCommons
}

\begin{document}
\maketitle

\begin{abstract}
AI safety evaluations often involve sensitive prompts and proprietary models. Evaluators want to keep their evaluation prompts secret from model owners to prevent purposeful or incidental data leakage, ensuring longer usefulness of their benchmark. Model owners in turn wish to keep their proprietary model weights and inference code within their own secure infrastructure to protect their IP. We present a way forward for ``double-blind evals'', which provide cryptographic guarantees that an evaluator's prompts are not shared with the model owner, while simultaneously maintaining the confidentiality of the model weights and inference code. To accomplish this, we use secure enclaves that encrypt data in RAM at the hardware level. In a world-first pilot, we test this approach by evaluating a proprietary Google DeepMind model (Gemini 2.5 Flash Lite) against a selection of the private AILuminate benchmark corpus, using an NVIDIA H100 secure enclave on Google Cloud Platform, and using OpenMined's PySyft to maintain data and model privacy. We believe this effort marks the first use of this technology to provide mutual secrecy for one organization's frontier-class proprietary model weights and another organization's closed benchmark.
\end{abstract}

\section{Introduction}

One of the core questions in the evaluation of modern AI systems is whether the public can trust the benchmarks. \citet{singh2025leaderboard} show that ``benchmark hacking'' is an urgent problem, with one frontier lab testing 27 private model variants on Chatbot-Arena, then publishing the top scorer.

However, accidental benchmark contamination is also a serious problem, and one that may not be solvable with trust, contracts, and zero-logging policies alone. \citet{xu2024benchmarking} show strong evidence of benchmark data leakage in both pre- and post-training for roughly half of the 31 models they tested. Furthermore, \citet{schaeffer2026contamination} show that test set contamination inflates measured benchmark performance, with the inflation growing with both the amount of contamination and model size.

To prevent both incidental and deliberate post-training on benchmark prompts, model evaluators can decide to keep their benchmark prompts completely private. For open-weight models, this is trivial---evaluators can simply download the weights and run their evaluations on infrastructure they control. However, for proprietary models, things are more complex. Although zero-logging protocols and rigorous contractual safeguards have long kept external test prompts confidential, they require trust between the parties that private evaluation data does not leak into future pre- or post-training runs. 

Evaluators cannot simply test proprietary models via the model owner's APIs, which expose benchmark prompts to the model owner in plain text and potentially compromise test set integrity. Even contractual arrangements and zero-logging systems could fail to prevent inadvertent leaks of private evaluation data into future pre- or post-training runs.

One alternative is for these private-benchmark evaluators to request a copy of proprietary model weights, so that they can evaluate the model themselves with confidence that they have not disclosed their benchmark to any other party. In fact, many evaluators require (and receive) direct access to model weights for certain high stakes safety/alignment/capability evaluations, including activation probing, representation steering, token-level log-likelihood audits, and deterministic offline verification.

However, most evaluators struggle to obtain copies of frontier weights because they represent massive capital investment, intellectual property, and potential dual-use capabilities that are important for model developers to safeguard. Consequently, only evaluators with deep relationships with model owners (e.g. evaluators working as employees of such organizations) tend to obtain that level of access. To the best of our knowledge, no external model evaluator has received a copy of frontier model weights in the clear for the purpose of protecting the integrity of a benchmark.

As a result, the model evaluators and model owners are stuck in a bind, in which evaluators cannot protect their benchmarks while evaluating proprietary frontier systems, and model owners cannot obtain the level of robust endorsement that a truly double-blind eval would provide (there is always the slight chance of leakage undermining a score).

\begin{figure}[t]
\centering
\resizebox{\textwidth}{!}{%
\begin{tikzpicture}[
    font=\sffamily,
    >=Stealth,
    partybox/.style={
        draw=#1, line width=1.4pt, fill=#1!6, rounded corners=6pt,
        inner sep=8pt, text width=4.6cm, align=center
    },
    innercard/.style={
        draw=#1, line width=1.1pt, fill=white, rounded corners=5pt,
        inner sep=7pt, text width=4.2cm, align=center
    },
    edgelabel/.style={
        font=\scriptsize, align=center, inner sep=2pt
    }
]

\node[partybox=gdmblue] (owner) at (-5.6, 6.4) {
    {\normalsize\textbf{Model Owner}}\\[1pt]
    \rule{4.2cm}{0.4pt}\\[4pt]
    \raggedright\scriptsize
    $\bullet$ Proprietary Weights ($W_{\text{model}}$)\\
    $\bullet$ Mock Model Interface $\widetilde{W}_{\text{model}}$\\
};

\node[partybox=gdmgreen] (evaluator) at (5.6, 6.4) {
    {\normalsize\textbf{Benchmark Evaluator}}\\[1pt]
    \rule{4.2cm}{0.4pt}\\[4pt]
    \raggedright\scriptsize
    $\bullet$ Private Benchmark Prompts ($D_{\text{eval}}$)\\
    $\bullet$ Evaluation Script \& Policies\\
};

\draw[->, line width=1.2pt, draw=darkgray] (-3.0, 6.9) -- node[above=2pt, edgelabel] {\textbf{(1)} Publish Mock Interface $\widetilde{W}_{\text{model}}$} (3.0, 6.9);
\draw[<->, line width=1.2pt, draw=darkgray] (3.0, 5.8) -- node[below=2pt, edgelabel] {\textbf{(2)} Submit Evaluation Code \& Policies} (-3.0, 5.8);

\node[draw=openminedpurple, dashed, line width=1.1pt, fill=attestbg, rounded corners=6pt, inner sep=7pt, text width=5.6cm, align=center] (attest) at (0, 3.6) {
    {\normalsize\textbf{\mbox{(3) Hardware Remote Attestation}}}\\[1pt]
    {\scriptsize\textbf{(IETF RATS / Silicon Root Keys)}}\\[2pt]
    \textmd{\tiny Enclave identity verified independently by both parties}
};

\draw[->, dashed, line width=1.1pt, draw=openminedpurple] (0, 1.4) -- node[pos=0.5, right=3pt, font=\tiny\bfseries, text=openminedpurple] {Attestation Quote} (0, 2.9);

\draw[->, dashed, line width=1.1pt, draw=openminedpurple] (attest.west) -| (-4.8, 4.8);
\draw[->, dashed, line width=1.1pt, draw=openminedpurple] (attest.east) -| (4.2, 4.8);

\node[draw=openminedpurple, line width=1.6pt, fill=enclavegray, rounded corners=8pt, inner sep=12pt, text width=16.8cm, align=center] (enclave) at (0, -1.1) {
    {\normalsize\textbf{Google Cloud Confidential Space Secure Enclave}}\\[1pt]
    {\scriptsize\color{darkgray}NVIDIA H100 GPU + Intel TDX Host \ \textbar\ \ Ephemeral Lifecycle (Zero State Remanence)}\\[10pt]

    \begin{tikzpicture}
        \node[innercard=gdmblue] (host) at (-4.7, 0.5) {
            {\footnotesize\textbf{Host RAM (Intel TDX)}}\\[2pt]
            \raggedright\tiny
            $\bullet$ Hardware-encrypted RAM (AES-256)\\
            $\bullet$ Attested PySyft runtime container\\
            $\bullet$ Private evaluation prompts $D_{\text{eval}}$\\
        };

        \node[innercard=openminedpurple] (gpu) at (4.7, 0.5) {
            {\footnotesize\textbf{Confidential GPU (H100)}}\\[2pt]
            {\raggedright
            \tiny
            $\bullet$ Hardware-encrypted HBM3 VRAM\\
            $\bullet$ On-die AES-XTS-256 crypto engines\\
            $\bullet$ Proprietary weights $W_{\text{model}}$ in VRAM\\}
        };

        \draw[<->, line width=1.3pt, draw=darkgray] (host.east) --
            node[above=2pt, font=\scriptsize\bfseries] {Encrypted PCIe DMA}
            node[below=2pt, font=\tiny\color{darkgray}] {Hardware Firewalls}
            (gpu.west);

        \node[draw=darkgray!60, fill=stepbg, rounded corners=4pt, line width=0.7pt, text width=15.6cm, align=center, inner sep=6pt, font=\scriptsize] at (0, -1.4) {
            \textbf{Isolated Execution:} $\mathcal{R} \leftarrow \text{EvalFn}(W_{\text{model}}, D_{\text{eval}})$ under Structured Output Privacy Policies
        };
    \end{tikzpicture}
};

\draw[->, line width=1.4pt, draw=gdmblue] (-6.6, 4.8) -- node[pos=0.5, left=3pt, edgelabel, text=gdmblue!90!black] {\textbf{(4a) Stream $W_{\text{model}}$}\\[1pt]{\tiny (Encrypted mTLS into VRAM)}} (-6.6, 1.4);

\draw[->, line width=1.8pt, draw=gdmgreen] (5.4, 1.4) -- node[pos=0.5, right=3pt, edgelabel, text=gdmgreen!90!black] {\textbf{(5) Return Results $\mathcal{R}$}\\[1pt]{\tiny (Bounded metrics)}} (5.4, 4.8);

\draw[->, line width=1.4pt, draw=gdmgreen] (8.6, 4.8) -- node[pos=0.5, right=3pt, edgelabel, text=gdmgreen!90!black] {\textbf{(4b) Stream $D_{\text{eval}}$}\\[1pt]{\tiny (Encrypted mTLS into RAM)}} (8.6, 1.4);

\end{tikzpicture}%
}
\caption{Double-Blind Evaluation (DBE) Framework. (1) The Model Owner publishes a mock interface to the model. (2) The Benchmark Evaluator drafts private benchmark prompts and code using the mock interface. (3) Both parties cryptographically verify remote hardware attestation. (4) Model weights and private benchmark prompts are streamed over encrypted TLS tunnels directly into a hardware-encrypted enclave. (5) The benchmarks are evaluated in this isolated environment and aggregate metrics are sent to the evaluator, and finally the enclave is decommissioned.}
\label{fig:dbe_architecture}
\end{figure}
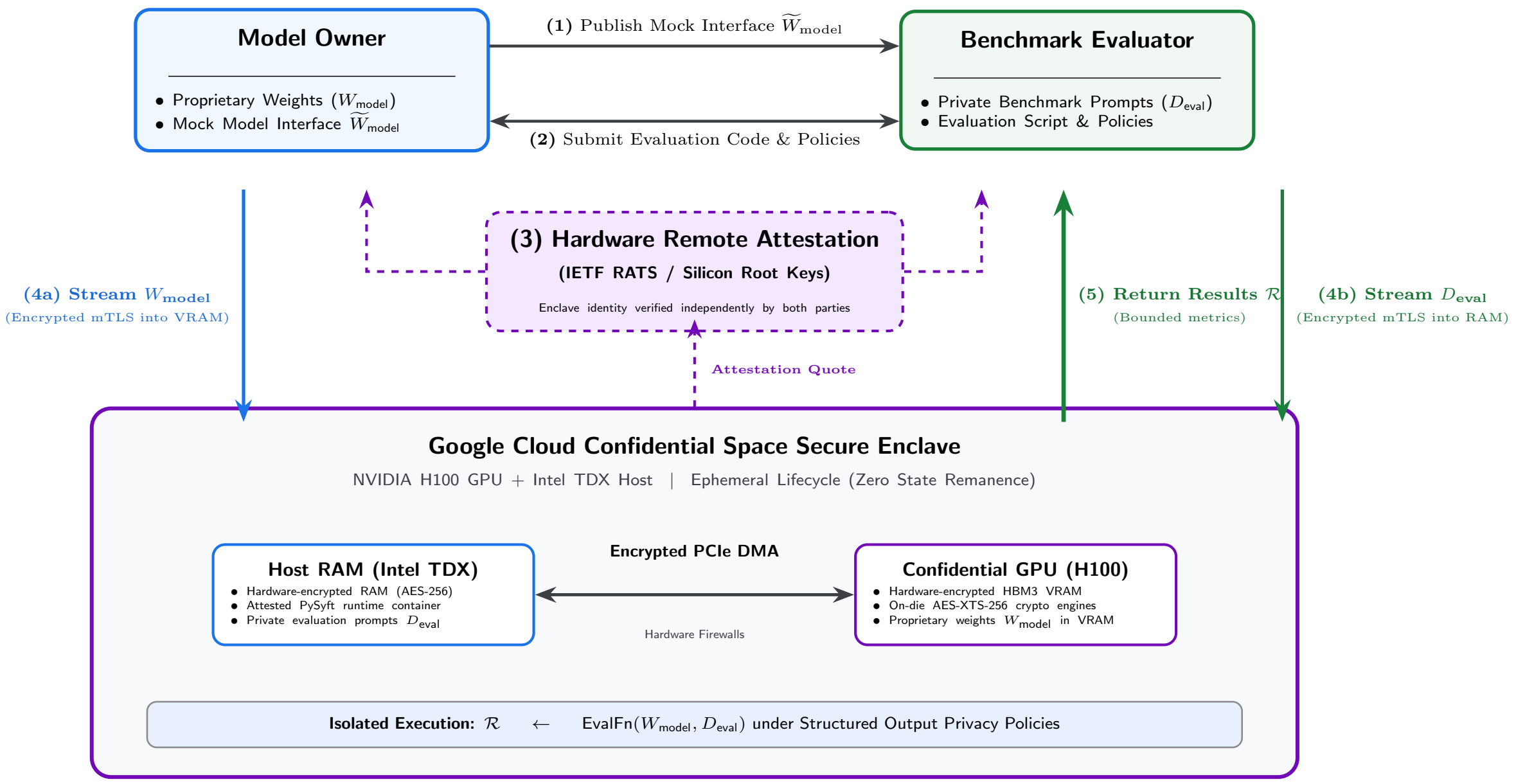

As a remedy, we present the Double-Blind Evaluation (DBE) framework (Figure~\ref{fig:dbe_architecture}), a cryptographic and hardware-enforced evaluation environment that provides mutual secrecy. This framework prevents even inadvertent data leakage from evaluations into future training runs by providing a temporary, cryptographically secure environment that hosts both evaluations and proprietary models. DBE provides this mutual secrecy through a PySyft Datasite architecture that executes evaluations on a two-tier hardware enclave hierarchy on Google Cloud Platform (GCP) Confidential Space (a3-highgpu-1g instances), operating with less than 5\% compute overhead \citep{nvidia2023confidential}.

The success of this framework depends on whether it inspires sufficient confidence across separate organizations such that they allow their proprietary assets to be protected through it. In 2024, OpenMined successfully pilot tested the foundational PySyft Datasite architecture and GPU enclave coordination model in a small-scale proof-of-concept with the UK AI Safety Institute (AISI) and Anthropic (GPT-2 evaluated against a 5-row sample of CAMEL-bio, an AISI hosted dataset; \citealp{openmined2024secureenclaves}). This pilot described the remaining security features needed to inspire organizational confidence. But it did not accomplish the key objective above: protection of two organizations' proprietary assets. 

More recently, MLCommons' MedPerf platform has federated benchmark evaluation in medical AI to assess private data that never leaves the holding institution \citep{karargyris2023medperf}, and evaluated encrypted medical imaging models inside a Google Cloud Confidential Space \citep{kolga2026medperf}.

Below we provide technical details describing our current implementation and report on the first live deployment of DBE for a proprietary model evaluated on a closed benchmark. With the AI Verification and Evaluation Research Institute (AVERI) and MLCommons, we evaluated Google DeepMind's Gemini 2.5 Flash Lite using private evaluations from the MLCommons AILuminate (AIRR 1.0) benchmark corpus inside an isolated GCP-hosted NVIDIA H100 secure enclave. With Singapore AISI, we evaluated Google DeepMind's Gemini 2.5 Flash Lite using a private prompt set focused on harmful content elicitation in Singapore's context.

\section{Double-blind Eval Mechanics}

At a high level, a secure enclave wherein compute is cryptographically protected doesn't provide double-blind capability out of the box. Instead, it provides two low level guarantees upon which a double-blind capability can be constructed. To describe the scheme, we first describe the low level guarantees provided by an enclave, and then describe the procedure through which we construct the higher level DBE capability.

\subsection{High-level Enclave Guarantees}

The problem that a CPU or GPU enclave is designed to address is the (theoretical) problem of an untrustworthy compute provider. In this theoretical scenario, a compute provider can violate two types of principles: privacy and veracity \citep{trask2020structured}.

\begin{description}
\item[Privacy:] In theory, a compute provider can see all data operating within its computers.
\item[Veracity:] In theory, a compute provider can pretend to execute a program/data a user has requested while actually running some other program/data, sending false results back to the customer.
\end{description}

A CPU/GPU enclave solves this problem not by making privacy or veracity impossible to compromise outright (i.e. a ``trustless system''), but by creating a scenario that's close thereto---wherein a compute provider and a hardware manufacturer would have to collude in order to compromise those guarantees, because the system's encryption works at the hardware level. Consequently, if a compute user believes that a compute provider and hardware manufacturer are unlikely to collude against them, then a compute user believes that a CPU/GPU enclave provides strong privacy and veracity guarantees.

\subsection{How an Enclave Provides Privacy and Veracity Guarantees}

A CPU/GPU enclave (such as Intel TDX or AMD SEV-SNP) is a normal computing chip modified by the hardware manufacturer to hold unique secret numbers, burned into the chip when it is manufactured. From these secrets the chip derives a private key (with a corresponding public key), and this is the foundation of an enclave's root of trust: the manufacturer publishes certificates vouching for each chip's public key, so anyone can check that a signature really came from a genuine chip. If these secrets ever leak beyond the chip or the manufacturer mishandles the certification keys it uses to vouch for them, the root of trust is broken. However, so long as they remain unknown to everyone else (especially the cloud provider), they form a powerful foundation for privacy and veracity guarantees.

With respect to privacy, the enclave ensures that a cloud provider cannot see into its operations. To do this, the chip generates a separate, temporary memory-encryption key, different from the private key above, created fresh each time an enclave starts and destroyed when it stops and uses it to automatically encrypt everything the enclave holds in RAM. Data written to disk is encrypted as well, by software running inside the enclave, using keys that only ever exist inside that protected memory. Consequently, even if a cloud provider physically inspected the RAM or the hard disk, it could not recover the data in plaintext, because the keys live in dedicated hardware that no software, not even the cloud provider's can read out \citep{amd2020sevsnp,young2023snp}.

With respect to veracity, the enclave uses its private key to produce a signed hash of all of the software running on it at any given time, after which the enclave shares the signed hash externally such that outside parties can be confident in how their data will be handled if sent to the enclave (i.e. what the enclave's software will do with their data).

However, upon close inspection one might notice that this last veracity claim depends heavily on whether the software running inside of an enclave doesn't itself violate privacy or veracity. For example, if the enclave is running an SSH server and the firewall is disabled, an administrator could login and start changing the code after a signed hash has been disclosed to an outside party. Or to violate privacy, if the enclave is running black box code which an outside party doesn't trust, the enclave might take all of a user's data and copy it outside of the enclave. Given this risk, how might an enclave actually provide a privacy and veracity guarantee in practice?

For this, an enclave also needs a software stack which is available to the outside party to trust (e.g. open source or otherwise trusted), and that software stack needs to avoid privacy and veracity violations. Consequently, while an enclave might run what appears to be a normal bios, operating system, Python libraries, and other software, it often requires slightly modified versions of that software. For example, it requires an operating system with keyboard, mouse, screen, and SSH disabled, and applications whose outbound messaging has been removed or sufficiently constrained so as to not violate the privacy and veracity goals of the enclave.

The full stack of trusted/modified software running inside the enclave---firmware, guest kernel, init system, container runtime, and the application image itself---constitutes the trusted computing base (TCB). If it possesses a vulnerability, any one of these layers can, in principle, subvert the confidentiality and integrity guarantees the enclave is meant to provide, so a user must be able to establish that the enclave is running exactly the TCB they intend to trust, and nothing else. We achieve this with three mechanisms:

1. \textbf{Measurement}. Each layer of the TCB is cryptographically measured (hashed) as it is loaded, and the resulting measurements are included in an attestation report signed by a key rooted in the CPU vendor's hardware root of trust. The signature chains back to a vendor-issued certificate, so the report cannot be forged by the host, the cloud operator, or the enclave owner. Because each layer measures the next before transferring control to it, the chain of measurements covers the whole stack. The enclave additionally generates an ephemeral keypair at boot and embeds the public key in the report, so that the secure channel a user establishes is cryptographically bound to the attested TCB rather than merely to a machine that can present a valid report.

2. \textbf{Reproducible builds}. Every layer of the TCB is built deterministically from public source, so that any third party can rebuild it and obtain a bit-identical artifact, and hence the same measurement. This is what turns a measurement from an opaque number into a meaningful one: without reproducibility, a user can observe that the enclave is running some fixed software, but not which software.

3. \textbf{Verification}. Before releasing any data to the enclave, each user independently verifies the attestation report: they check the vendor signature chain, check that the report is fresh by requiring it to be bound to a user-supplied nonce, and compare the reported measurements against the expected measurements for the TCB they intend to trust. Data is released only if every measurement matches. Users may derive those expected values themselves by rebuilding the TCB from source, or delegate the rebuild to any party they trust to do it---an auditor, a peer participant, or an independent rebuilder---since reproducibility means every honest rebuilder arrives at the same values, and a single one who rebuilds and publishes a mismatch is enough to expose a substituted TCB. Verification itself is performed by each participant; no party has to trust another participant's verdict, nor a central verification service.

The lowest layers of the TCB---CPU microcode and secure-processor firmware---are closed-source and cannot be rebuilt, so these we trust the CPU vendor for: AMD on SEV-SNP, Intel on TDX. That is unavoidable and already implied: the same vendor's root of trust signs the attestation in the first place. It is also constrained: their security versions are bound into the report's signing key, so a platform cannot misreport what it runs, and a verifier can reject any version below a chosen floor. Crucially, this layer is identical for every user of the hardware and cannot be varied per deployment, so it offers an operator no way to target a specific victim. Reproducibility applies to everything above it, where the deployment-specific software lives.

But when an enclave is running an open source (or otherwise trusted) software stack with privacy/veracity violating features disabled, and when that enclave uses its private key to sign that software being run and to encrypt all data written to RAM or to disk, that CPU/GPU enclave can provide an extraordinary guarantee to the outside user. If the outside user trusts that their cloud provider and hardware manufacturer have insufficient incentives to collude against them, they can have very high confidence that their application data is seen by no-one, and that they are running the code they believe they are running.

\subsection{From Enclaves to Theoretical SMPC Capability}

However, protection from a cloud provider does not itself provide the guarantees necessary for a double blind evaluation, which requires privacy and veracity guarantees between two users of a VM (as opposed to privacy and veracity guarantees between a cloud and a single user).

However, we can build from the former to the latter. First, if two users of a VM both receive a signed hash of the software running in an enclave, and if they both consent to how their data will be used by that software, then they can safely send their data into the enclave, wait for the enclave to perform that computation, and consume the outputs.

The classic example here is called the ``Millionaire's Problem'', wherein two (theoretical) millionaires want to know who has more money but they don't want to disclose their exact bank balance to each other. So, to calculate the simple threshold function, one of them could launch an enclave which is running open source software which expects a message from either of them containing their bank balance, and which then outputs who has more money, deletes everything, and shuts down. One of the two millionaires would then launch the enclave with this software, they would both check the signed hash to ensure it's running the open source software they expect, and then upload their bank balances and wait for the result. And in this way, they can leverage an enclave to perform secure multi-party computation (SMPC) between them.

\subsection{From Theoretical SMPC to Double-blind Eval}

However, even this flavor of SMPC isn't sufficient to perform a double-blind evaluation in practice. Because while in the Millionaire's Problem all parties could co-agree upon all code which would be run (by looking at the code), and they only provided state, in the context of a double-blind evaluation a model owner needs to provide inference code and an evaluator often needs to provide evaluation and analysis code. And in many cases, each of these respective codebases contains IP or trade secrets which ought not be disclosed to the other party. This undermines one of the requirements above, that all parties view the code before it is run in the enclave to ensure that it adheres to their policies. If each party must agree to unknown code submitted by the other party, how can either be sure that privacy or veracity is maintained?

For this, a unique assumption provides a bridge. While each party's codebase might contain secrets which need to be hidden, that codebase is composed of smaller functions which themselves are composed of even smaller functions. Continuing downward, at some point the units of computation are so elementary as to no longer contain any proprietary IP. And in this realization, a privacy strategy emerges.

If hidden code can be guaranteed to only run methods which themselves do not send any data out of the enclave (or write data to a location which would send data out of the enclave), then the hidden code does not send data out of the enclave and is upholding the privacy guarantee.

Concretely, if a model owner is submitting inference code written entirely in libraries like Jax or PyTorch, then all of their model IP is a series of deep learning framework method calls which themselves are open source, and which have clear boundaries between network-bound activities (e.g. tensorflow server) and non-network bound local computation (e.g. \texttt{th.Tensor.add}). Consequently, if a model owner can reveal most of their code to the other party, but feels the need to redact portions, they can do so in a way where the enclave guarantees that the redacted code only calls methods from a specific allow-list of non-network-bound methods from the associating deep learning framework. Taken together, the AI evaluator can obtain a high degree of confidence that the redacted inference code of an AI model is not able to expatriate their benchmark or evaluation methodology.

Inversely, if an AI evaluator wishes to hide some of their evaluation methodology from the model owner, they too can submit their code with redactions as long as the redacted code can exclusively leverage methods from an allowlist of non-ip-leaking but non-network-calling methods. Taken together, both parties can bring mutually secret assets (model, benchmark) and code (inference code, and eval code), hide that asset from the other party, while learning sufficient information about their counterparty's hidden asset to be confident of the overall privacy and veracity of the evaluation. And it is this submission and approval of code which libraries like PySyft facilitate, and allow-list verified redaction which libraries like \texttt{syft-restrict} provide during the submission process.

\subsection{A Double-blind Eval in Practice with PySyft}

Altogether, when properly used, PySyft can facilitate this chain of guarantees from a GPU enclave's root of trust up to high level privacy and veracity guarantees between a model owner and an evaluator through the following steps.

\begin{enumerate}
\item To begin a double blind eval, a model owner gathers together the weights and inference code of a model they wish to make available for evaluation, and an external evaluator gathers together evaluation samples and evaluation code. They align and confirm their intent to collaborate.
\item Upon establishing intent, one party elects to launch (and pay for) a GPU enclave, perhaps within a cloud provider like GCP. This party obtains one special privilege as the enclave hosting party: the ability to turn off the enclave. But the enclave-deploying party obtains no other special privileges relative to the non-enclave-deploying party. The enclave is launched with an enclave compatible bios, drivers, operating system, and PySyft docker container.
\item Upon launch, each party opens a python runtime (e.g. Google Colab) and leverages the PySyft client to request that the enclave produce a signed hash of its software (a certificate).
\item (Optional) If each party is going to need to interface their code tightly with the other party's code, each can use the PySyft client to submit a fake/mock version of their model or benchmark into the enclave, which the other party can download and use as a test harness when developing the final version of their code. Alternatively, when APIs are standard (e.g. basic LLM inference), one party can simply indicate to the other which standard APIs they are using.
\item Using the PySyft python client, each party uploads their (possibly redacted) code and private assets into the enclave (model, benchmark, etc.). Code is only run when approved by both parties.
\item Each party receives the redacted code from the enclave (with attestation from \texttt{syft-restrict} that the redactions only contain method calls from an allowlist). They then use the PySyft client to review and approve the code, messaging the enclave as such.
\item Once the enclave has received approval for the computation from each party, it executes the code and releases the result. During submission of the code, the parties can choose who receives the result (both or one party).
\end{enumerate}

Such is the high level process we propose for a double blind evaluation.

\section{Empirical Demonstration: Evaluating Gemini 2.5 Flash Lite using AILuminate}

We demonstrate DBE end-to-end using closed benchmark prompts from MLCommons and a proprietary Google DeepMind model, in partnership with AVERI.

\subsection{Experimental Setup}

We evaluated Google DeepMind's Gemini 2.5 Flash Lite served via Google's Model Server operating over Unix Domain Sockets within the Google Cloud confidential space. During initialization, uncompressed model weights were streamed into the enclave alongside DAT verification keys and security realm configurations.

The benchmark used in this experiment was drawn from the reserve set (i.e., prompts that have never been processed by any model) of AILuminate. The prompts covered critical hazard domains including Chemical, Biological, Radiological, Nuclear, and Explosive (CBRNE) hazards, cyberattacks, hate speech, self-harm, and violent crime elicitation. The hazards and violation criteria for AILuminate AIRR 1.4 were produced within MLCommons AI Risk and Reliability (AIRR) working groups, including representatives from AVERI. The prompts were produced under strict non-disclosure and origination requirements with data providers and not shared broadly with members of the working groups. The prompts and outputs used in the experiment were encrypted and decrypted by AVERI and the outputs were evaluated by AVERI staff.

We used a Google Cloud Platform (GCP) A3 Confidential VM (a3-highgpu-1g instance with Intel Trust Domain Extensions (TDX) host memory encryption). This was equipped with an NVIDIA H100 80GB Confidential GPU, orchestrating an attested software stack comprising OpenMined PySyft v0.10.x, Google's Gemini runtime library stack, and the NVIDIA Attestation SDK.

\section{Discussion and Future Directions}

This deployment of Double-Blind Evaluation constitutes the first instance where model benchmark developers have successfully evaluated a frontier class proprietary model, minimizing the danger of compromising trade secrets or benchmark integrity with cryptographic guarantees. Independent evaluation agencies in government such as national AI Safety and Security Institutes as well as non-profit organizations such as AVERI can require access to frontier model weights in order to conduct evaluations. DBE provides a framework for rigorous safety and capability evaluation without compromising the secrecy of non-public benchmarks, nor requiring model developers to transfer weights to an evaluator's infrastructure.

As observed in initial multi-stakeholder enclave pilots \citep{openmined2024secureenclaves}, the primary bottleneck to secure double blind evaluation is no longer hardware compute overhead. It is instead the procedural overhead and human coordination required for legal agreements and code review.

While these initial efforts required a high level of joint human coordination, the long run goal should be standardized cryptographic attestation pipelines with trust guarantees that abstract away complex dependency hashes and keys with near zero human overhead. The eventual goal for the industry would be an analogy to the visual ``HTTPS lock icon'' on the web.

An additional challenge involved running Gemini 2.5 Flash Lite using only layers present in open source libraries. It was deemed to be too significant of an engineering challenge for this project to eliminate proprietary method implementations, such that not all code could be inspected or allowlisted. AVERI was informed of this challenge and accepted the overall setup with the enclave. Additionally, although reference values are published for the Confidential Space guest OS, its source is open, and its build pipeline is externally validated, the individual builds are not independently reproducible, as they take private signing keys as inputs. Furthermore, we rely on Google's services to sign and verify the attestation report, which places Google in the verification path and thus increases the trust placed in it. Future work will revisit these security improvements and inference implementation tasks.

As a next step, we will explore scaling this approach. Frontier models are growing beyond trillions of parameters, and so evaluation efforts will need distributed confidential clusters. Expanding DBE to many-node NVIDIA H100/B200 compute clusters with encrypted connections alongside deferred execution frameworks that mask internal model architecture represents the next milestone for confidential AI evaluation.

Looking even further out, we speculate that the verification and attestation assurances could prove useful in securely evaluating highly advanced models on capabilities like cybersecurity. For example, if a model were to try to adjust its harness, weights, or sandbox, the enclave hash would no longer match attestation and the instance would immediately come to a halt.

\section{Conclusion}

Double-Blind Evaluation (DBE) provides a cryptographically secure and verifiable solution to keep benchmarks confidential to evaluators and weights confidential to model developers. By combining PySyft's Structured Transparency policies with hardware-level memory encryption, DBE provides cryptographic guarantees that protect the confidentiality of each party's assets.

We validate this paradigm through the first live deployment evaluating Google DeepMind's Gemini 2.5 Flash Lite against the private MLCommons AILuminate benchmark in partnership with AVERI, and against a private prompt set focused on harmful content elicitation in Singapore's context in partnership with Singapore AISI. DBE establishes a way forward for independent secure nonpublic evaluations on proprietary systems.

\section*{Acknowledgments}

The authors thank the engineering teams at OpenMined, Google DeepMind, Google Cloud, AVERI, and MLCommons for their technical contributions and support throughout this collaboration. We also want to thank Helen King, Tim Dierks, Rohin Shah, Wei Huang, Owen Larter and Tom Lue at Google DeepMind for their leadership.

\bibliographystyle{apalike}
\bibliography{references}

\begin{thebibliography}{}

\bibitem[{Advanced Micro Devices, Inc.}, 2020]{amd2020sevsnp}
{Advanced Micro Devices, Inc.} (2020).
\newblock {AMD} {SEV-SNP}: Strengthening {VM} isolation with integrity protection and more.
\newblock AMD White Paper.

\bibitem[Apsey et~al., 2023]{nvidia2023confidential}
Apsey, E., Rogers, P., O'Connor, M., and Nertney, R. (2023).
\newblock Confidential computing on {NVIDIA} {H100} {GPUs} for secure and trustworthy {AI}.
\newblock NVIDIA Technical Blog.
\newblock \url{https://developer.nvidia.com/blog/confidential-computing-on-h100-gpus-for-secure-and-trustworthy-ai/}.

\bibitem[Karargyris et~al., 2023]{karargyris2023medperf}
Karargyris, A., Umeton, R., Sheller, M.~J., et~al. (2023).
\newblock Federated benchmarking of medical artificial intelligence with {MedPerf}.
\newblock {\em Nature Machine Intelligence}, 5:799--810.

\bibitem[Kolga and Mattson, 2026]{kolga2026medperf}
Kolga, R. and Mattson, P. (2026).
\newblock Advancing brain tumor research with privacy-first {AI}.
\newblock Google Cloud Blog.
\newblock \url{https://cloud.google.com/blog/products/identity-security/privacy-first-medical-ai-with-medperf-and-google-cloud}.

\bibitem[Schaeffer et~al., 2026]{schaeffer2026contamination}
Schaeffer, R., Kazdan, J., Abbasi, B., Liu, K.~Z., Miranda, B., Ahmed, A., Barez, F., Puri, A., Biderman, S., Mireshghallah, N., et~al. (2026).
\newblock Quantifying the effect of test set contamination on generative evaluations.
\newblock {\em arXiv preprint arXiv:2601.04301}.

\bibitem[Singh et~al., 2025]{singh2025leaderboard}
Singh, S., Nan, Y., Wang, A., Dsouza, D., Kapoor, S., {\"U}st{\"u}n, A., Koyejo, S., Deng, Y., Longpre, S., Smith, N.~A., et~al. (2025).
\newblock The leaderboard illusion.
\newblock In {\em Advances in Neural Information Processing Systems}, volume~38.
\newblock Datasets and Benchmarks Track. arXiv:2504.20879.

\bibitem[Trask et~al., 2020]{trask2020structured}
Trask, A., Bluemke, E., Collins, T., Garfinkel, B., Drexler, E., {Ghezzou Cuervas-Mons}, C., Gabriel, I., Dafoe, A., and Isaac, W. (2020).
\newblock Beyond privacy trade-offs with structured transparency.
\newblock {\em arXiv preprint arXiv:2012.08347}.

\bibitem[Trask et~al., 2024]{openmined2024secureenclaves}
Trask, A., Yesilyurt, A.~B., Farkas, B., Ezenwaka, C., Popa, C., Buckley, D., van~der Wel, E., Mosconi, F., Han, G., Junior, I., et~al. (2024).
\newblock Secure enclaves for {AI} evaluation.
\newblock \url{https://openmined.org/blog/secure-enclaves-for-ai-evaluation/}.
\newblock OpenMined Technical Blog, in partnership with UK AISI and Anthropic.

\bibitem[Xu et~al., 2024]{xu2024benchmarking}
Xu, R., Wang, Z., Fan, R.-Z., and Liu, P. (2024).
\newblock Benchmarking benchmark leakage in large language models.
\newblock {\em arXiv preprint arXiv:2404.18824}.

\bibitem[Young, 2023]{young2023snp}
Young, J. (2023).
\newblock Oh {SNP}! {VMs} get even more confidential.
\newblock Google Cloud Blog.
\newblock \url{https://cloud.google.com/blog/products/identity-security/rsa-snp-vm-more-confidential}.

\end{thebibliography}

\end{document}